\documentclass[runningheads,a4paper]{llncs}

\usepackage{amssymb}
\usepackage{graphicx}

\usepackage{enumitem}
\usepackage{color}
\usepackage{algpseudocode}

\usepackage{array}
\newcolumntype{L}[1]{>{\raggedright\let\newline\\\arraybackslash\hspace{0pt}}m{#1}}
\newcolumntype{C}[1]{>{\centering\let\newline\\\arraybackslash\hspace{0pt}}m{#1}}
\newcolumntype{R}[1]{>{\raggedleft\let\newline\\\arraybackslash\hspace{0pt}}m{#1}}

\usepackage{url}

\urldef{\mailsa}\path|{nqminh, giovanni, goran_topic, aizawa}@nii.ac.jp|    
\newcommand{\keywords}[1]{\par\addvspace\baselineskip
\noindent\keywordname\enspace\ignorespaces#1}

\usepackage{array}
\newcolumntype{L}[1]{>{\raggedright\let\newline\\\arraybackslash\hspace{0pt}}m{#1}}
\newcolumntype{C}[1]{>{\centering\let\newline\\\arraybackslash\hspace{0pt}}m{#1}}
\newcolumntype{R}[1]{>{\raggedleft\let\newline\\\arraybackslash\hspace{0pt}}m{#1}}

\begin{document}

\mainmatter  

\title{Which one is better: presentation-based or content-based math search?}

\titlerunning{Which one is better: presentation-based or content-based math search?}

%
%

\author{Minh-Quoc Nghiem$^{~1,4}$, Giovanni Yoko Kristianto$^{~2}$,\\ Goran Topi\'{c}$^{~3}$, Akiko Aizawa$^{~2,3}$}

\authorrunning{M.Q. Nghiem, G.Y. Kristianto, G. Topi\'{c}, A. Aizawa}


\institute{$^{1~}$Ho Chi Minh City University of Science,\\
$^{2~}$The University of Tokyo,\\
$^{3~}$National Institute of Informatics,\\
$^{4~}$The Graduate University for Advanced Studies,\\
\mailsa\\
}

%
%

\toctitle{Lecture Notes in Computer Science}
\tocauthor{Authors' Instructions}
\maketitle

\begin{abstract}
Mathematical content is a valuable information source and retrieving this content has become an important issue. 
This paper compares two searching strategies for math expressions: presentation-based and content-based approaches. 
Presentation-based search uses state-of-the-art math search system while content-based search uses semantic enrichment of math expressions to convert math expressions into their content forms and searching is done using these content-based expressions. 
By considering the meaning of math expressions, the quality of search system is improved over presentation-based systems.
\end{abstract}

\section{Introduction}

The issue of retrieving mathematical content has received considerable critical attention~\cite{2013-NTCIR-Aizawa}.
Mathematical content is a valuable information source for many users and is increasingly available on the Web.
Retrieving this content  is becoming more and more important.

Conventional search engines, however, do not provide a direct search mechanism for mathematical expressions.
Although these search engines are useful to search for mathematical content, these search engines treat mathematical expressions as keywords and fail to recognize the special mathematical symbols and constructs.
As such, mathematical content retrieval remains an open issue.

Some recent studies have proposed mathematical retrieval systems based on the structural similarity of mathematical expressions~\cite{URL-DLMF,URL-SpringerLaTeXSearch,2005-DLMF-Youssef,2008-Query-Altamimi,2007-Query-Youssef,2007-Query-Miner}.
However, in these studies, the semantics of mathematical expressions is still not considered.
Because mathematical expressions follow highly abstract and also rewritable representations, structural similarity alone is insufficient as a metric for semantic similarity.

Other studies~\cite{URL-Wolfram,2008-MathGO-Adeel,2006-MWS-Kohlhase,2009-MathDa-Yokoi,2013-NTCIR-Kohlhase,2012-CIKM-Nguyen} have addressed semantic similarity of mathematical formulae, but this required content-based mathematical formats such as content MathML~\cite{2010-MathML} and OpenMath~\cite{2004-OpenMath}.
Because almost all mathematical content available on the Web is presentation-based, these studies used two freely available toolkits, SnuggleTeX~\cite{URL-SnuggleTeX} and LaTeXML~\cite{URL-LaTeXML}, for semantic enrichment of mathematical expressions.
However, much uncertainty remains about the relation between the performance of mathematical search system and the performance of the semantic enrichment component.

Based on the observation that mathematical expressions have meanings hidden in their representation, the primary goal of this paper is making use of mathematical expressions' semantics for mathematical search.
To accomplish this problem of retrieving semantically similar mathematical expressions, we use the results of state-of-the-art semantic enrichment methods.
This paper seeks the answers to two questions.
\begin{itemize}
\item What is the contribution of semantic enrichment of mathematical expressions to content-based mathematical search systems?
\item Which one is better: presentation-based or content-based mathematical search?
\end{itemize}

To implement a {\it mathematical search system}, various challenges must be overcome.
First, in contrast to text which is linear, mathematical expressions are hierarchical: operators have different priorities, and expressions can be nested.
The similarity between two mathematical expressions is decided first by their structure and then by the
symbols they contain~\cite{2009-DML,DBLP-KamaliT13}.
Therefore, current text retrieval techniques cannot be applied to mathematical expressions because they only consider whether an object includes certain words.
Second, mathematical expressions have their own meanings.
These meanings can be encoded using special markup languages such as Content MathML or OpenMath.
A few existing mathematical search systems also make use of this information.
Such markup, however, is rarely used to publish mathematical knowledge related to the Web~\cite{2009-DML}.
As a result, we were only able to use presentation-based markup, such as Presentation MathML or \TeX{}, for mathematical expressions.

This paper presents an approach to a {\it content-based mathematical search system} that uses the information from {\it semantic enrichment of mathematical expressions} system.
To address the challenges described above, the proposed approach is described below.
First, the approach used Presentation MathML markup, a widely used markup for mathematical expressions.
This makes our approach more likely to be applicable in practice.
Second, a {\it semantic enrichment of mathematical expressions} system is used to convert mathematical expressions to Content MathML.
By getting the underlying semantic meanings of mathematical expressions, a {\it mathematical search system} is expected to yield better results.

The remainder of this paper is organized as follows.
Section 2 provides a brief overview of the background and related work.
Section 3 presents our method.
Section 4 describes the experimental setup and results.
Section 5 concludes the paper and points to avenues for future work.

\section{Mathematical Search System}

As the demand for mathematical searching increases, several mathematical retrieval systems have come into use~\cite{DBLP:journals/ijdar/ZanibbiB12}.
Most systems use the conventional text search techniques to develop a new mathematical search system~\cite{URL-DLMF,URL-SpringerLaTeXSearch}.
Some systems use specific format for mathematical content and queries~\cite{2005-DLMF-Youssef,2008-Query-Altamimi,2007-Query-Youssef,2007-Query-Miner,2009-MathDa-Yokoi}.
Based on the markup schema they use, current mathematical search systems are divisible into presentation-based and content-based systems.
Presentation-based systems deal with the presentation form whereas content-based systems deal with the meanings of mathematical formulae.

\subsection{Presentation-based systems}

\subsubsection{Springer LaTeXSearch}

Springer offers a free service, Springer \LaTeX{} Search~\cite{URL-SpringerLaTeXSearch}, to search for LaTeX code within scientific publications.
It enables users to locate and view equations containing specific \LaTeX{} code, or equations containing \LaTeX{} code that is similar to another \LaTeX{} string.
A similar search in Springer \LaTeX{} Search ranks the results by measuring the number of changes between a query and the retrieved formulae.
Each result contains the entire LaTeX string, a converted image of the equation, and information about and links to its source.



\subsubsection{MathDeX}

MathDeX (formerly MathFind~\cite{2006-Mathfind-Munavalli}) is a math-aware search engine under development by Design Science.
This work extends the capabilities of existing text search engines to search mathematical content.
The system analyzes expressions in MathML and decomposes the mathematical expression into a sequence of text-encoded math fragments.
Queries are also converted to sequences of text and the search is performed as a normal text search.

\subsubsection{Digital Library of Mathematical Functions}

The Digital Library of Mathematical functions (DLMF) project at NIST is a mathematical database available on the Web~\cite{URL-DLMF,2003-DLMF-Miller}.
Two approaches are used for searching for mathematical formulae in DLMF.
The first approach converts all mathematical content to a standard format.
The second approach exploits the ranking and hit-description methods.
These approaches enable simultaneous searching for normal text as well as mathematical content.

In the first approach~\cite{2005-DLMF-Youssef}, they propose a textual language, Textualization, Serialization and Normalization (TexSN). 
TeXSN is defined to normalize non-textual content of mathematical content to standard forms.
User queries are also converted to the TexSN language before processing.
Then, a search is performed to find the mathematical expressions that match the query exactly.
As a result, similar mathematical formulae are not retrieved.

In the second approach~\cite{2007-DLMF-Youssef}, the search system treats each mathematical expression as a document containing a set of mathematical terms.
The cited paper introduces new relevance ranking metrics and hit-description generation techniques.
It is reported that the new relevance metrics are far superior to the conventional tf-idf metric.
The new hit-descriptions are also more query-relevant and representative of the hit targets than conventional methods.






Other notable math search systems include Math Indexer and Searcher~\cite{MIaS2011}, EgoMath~\cite{EgoMath2011}, and ActiveMath~\cite{URL-ActiveMath}.

\subsection{Content-based systems}

\subsubsection{Wolfram Function}

The Wolfram Functions Site~\cite{URL-Wolfram} is the world's largest collection of mathematical formulae accessible on the Web.
Currently the site has 14 function categories containing more than three hundred thousand mathematical formulae.
This site allows users to search for mathematical formulae from its database.
The Wolfram Functions Site proposes similarity search methods based on MathML.
However, content-based search is only available with a number of predefined constants, operations, and function names.



\subsubsection{MathWebSearch}

The MathWebSearch system~\cite{2006-MWS-Kohlhase,2013-NTCIR-Kohlhase} is a content-based search engine for mathematical formulae.
It uses a term indexing technique derived from an automated theorem proving to index Content MathML formulae.
The system first converts all mathematical formulae to Content MathML markup and uses substitution-tree indexing to build the index.
The authors claim that search times are fast and unchanged by the increase in index size.

\subsubsection{MathGO!}

\cite{2008-MathGO-Adeel} proposed a mathematical search system called the MathGO! Search System.
The approach used conventional search systems using regular expressions to generate keywords.
For better retrieval, the system clustered mathematical formula content using K-Som, K-Means, and AHC.
They did experiments on a collection of 500 mathematical documents and achieved around 70--100 percent precision.

\subsubsection{MathDA}

Yokoi and Aizawa~\cite{2009-MathDa-Yokoi} proposed a similarity search method for mathematical expressions that is adapted specifically to the tree structures expressed by MathML.
They introduced a similarity measure based on Subpath Set and proposed a MathML conversion that is apt for it.
Their experiment results showed that the proposed scheme can provide a flexible system for searching for mathematical expressions on the Web.
However, the similarity calculation is the bottleneck of the search when the database size increases.
Another shortcoming of this approach is that the system only recognizes symbols and does not perceive the actual values or strings assigned to them.

\subsubsection{System of Nguyen et al.}

\cite{2012-CIKM-Nguyen} proposed a math-aware search engine that can handle both textual keywords and mathematical expressions.
They used Finite State Machine model for feature extraction, and representation framework captures the semantics of mathematical expressions.
For ranking, they used the passive--aggressive on-line learning binary classifier.
Evaluation was done using 31,288 mathematical questions and answers downloaded from Math Overflow~\cite{URL-MathOverflow}.
Experimental results showed that their proposed approach can perform better than baseline methods by 9\%.

\section{Methods}

\begin{figure*}[ht]
\centering
\includegraphics[width=10cm]{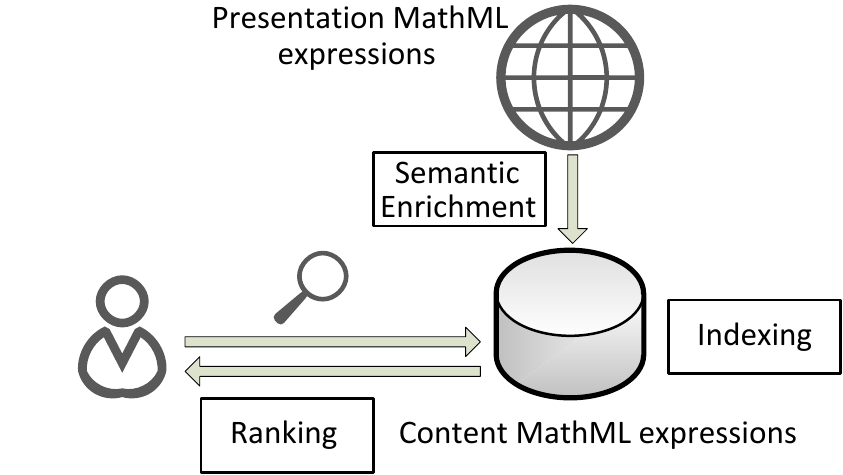}
\caption{System Framework.}
\label{fig:PIC501}
\end{figure*}

The framework of our system is shown in Fig. \ref{fig:PIC501}.
First, the system collects mathematical expressions from the web.
Then the mathematical expressions are converted to Content MathML using the {\it semantic enrichment of mathematical expressions} system of Nghiem et. al~\cite{2013-IEICE-Minh}.
Indexing and ranking the mathematical expressions are done using Apache Solr system~\cite{URL-Solr} following the method described in Topi\'{c} et. al~\cite{2013-NTCIR-Goran}.
When a user submits a query, the system also converts the query to Content MathML.
Then the system returns a ranked list of mathematical expressions corresponding to the user's queries.

\subsection{Data collection}

Performance analysis of a mathematical search system is not an easy task because few standard benchmark datasets exist, unlike other
more common information retrieval tasks.
Mathematical search systems normally build their own mathematical search dataset for evaluation by crawling and downloading mathematical content from the web.
Direct comparison of the proposed approach with other systems is also hard because they are either unavailable or inaccessible.

Recently, simpler and more rapid tests of mathematical search system have been developed.
The NTCIR-10 Math Pilot Task~\cite{2013-NTCIR-Aizawa} was the initial attempt to develop a common workbench for mathematical expressions search.
Currently, the NTCIR-10 dataset contains 100,000 papers and 35,000,000 mathematical expressions from ArXiv~\cite{URL-arXiv} which includes Content MathML markup.
The task was completed as an initial pilot task showing a clear interest in the mathematical search.
However, the Content MathML markup expressions are generated automatically using the LaTeXML toolkits. Therefore, this dataset is unsuitable to serve as the gold standard for the research described in the
present paper.

As Wolfram Functions Site~\cite{URL-Wolfram} is the only website that provides high-quality Content MathML markup for every expression, data for the search system was collected from this site.
The Wolfram Functions Site data have numerous attractive features, including both Presentation and Content MathML markups, and category for each mathematical expression.
In the experiment, the performance of {\it semantic enrichment of mathematical expressions} component will be compared directly with the system performance obtained using correct Content MathML expressions on Wolfram Functions Site data.

\subsection{Semantic enrichment of mathematical expressions}

The mathematical expressions were preprocessed according to the procedure described in Nghiem et. al~\cite{2013-IEICE-Minh}.
Given a set of training mathematical expressions in MathML parallel markup, rules of two types are extracted: segmentation rules and translation rules.
These rules are then used to convert mathematical expressions from their presentation to their content form.
Translation rules are used to translate (sub)trees of Presentation MathML markup to (sub)trees of Content MathML markup.
Segmentation rules are used to combine and reorder the (sub)trees to form a complete tree.

After using mathematical expression enrichment system to convert the expressions into content MathML, we use these converted expressions for indexing.
The conversion is not a perfect conversion, so there are terms that could not be converted.
The queries submitted to the search system are also processed using the same conversion procedure.

\subsection{Indexing}

The indexing step was prepared by adapting the procedure used by Topi\'{c} et. al~\cite{2013-NTCIR-Goran}.
This procedure used $pq$-gram-like indexing for Presentation MathML expressions.
We modified it for use with Content MathML expressions.
There are three fields used to encode the structure and contents of a mathematical expression: {\tt opaths}, {\tt upaths}, and {\tt sisters}.
Each expression is transformed into a sequence of keywords across several fields.
{\tt opaths} (ordered paths) field gathers the XML expression tree in vertical paths with preserved ordering.
{\tt upaths} (unordered paths) works the same as {\tt opaths} without the ordering information.
{\tt sisters} lists the sister nodes in each subtree.
Figure~\ref{fig5:encodingExample} presents an example of the terms used in the index of the expression $\sin(\frac{\pi}{8})$:$<apply><sin /><apply><times /><pi /><apply><power /><cn type=``integer">8</cn><cn type=``integer">-1</cn></apply></apply></apply>$.

\begin{figure*}[tbp]
  {\ttfamily \footnotesize
  \begin{description}[font=\textbf,itemsep=0mm]
    \item[opaths:]\hfill \\ 1\#1\#apply 1\#1\#1\#sin 1\#1\#2\#apply 1\#1\#2\#1\#times 1\#1\#2\#2\#pi
1\#1\#2\#3\#apply 1\#1\#2\#3\#1\#power 1\#1\#2\#3\#2\#cn\#8 1\#1\#2\#3\#3\#cn\#-1
    \item[opaths:]\hfill \\ 1\#apply 1\#1\#sin 1\#2\#apply 1\#2\#1\#times 1\#2\#2\#pi 1\#2\#3\#apply
1\#2\#3\#1\#power 1\#2\#3\#2\#cn\#8 1\#2\#3\#3\#cn\#-1
    \item[opaths:]\hfill \\  apply 1\#sin 2\#apply 2\#1\#times 2\#2\#pi 2\#3\#apply 2\#3\#1\#power
2\#3\#2\#cn\#8 2\#3\#3\#cn\#-1
    \item[opaths:] sin
    \item[opaths:] times
    \item[opaths:] pi
    \item[opaths:] apply 1\#power 2\#cn\#8 3\#cn\#-1
    \item[opaths:] power
    \item[opaths:] cn\#8
    \item[opaths:] cn\#-1
    \item[upaths:]\hfill \\  \#\#apply \#\#\#sin \#\#\#apply \#\#\#\#times \#\#\#\#pi \#\#\#\#apply \#\#\#\#\#power
\#\#\#\#\#cn\#8 \#\#\#\#\#cn\#-1
    \item[upaths:]\hfill \\  \#apply \#\#sin \#\#apply \#\#\#times \#\#\#pi \#\#\#apply \#\#\#\#power
\#\#\#\#cn\#8 \#\#\#\#cn\#-1
    \item[upaths:]\hfill \\  apply \#sin \#apply \#\#times \#\#pi \#\#apply \#\#\#power \#\#\#cn\#8 \#\#\#cn\#-1
    \item[upaths:] sin
    \item[upaths:] apply \#times \#pi \#apply \#\#power \#\#cn\#8 \#\#cn\#-1
    \item[upaths:] times
    \item[upaths:] pi
    \item[upaths:] apply \#power \#cn\#8 \#cn\#-1
    \item[upaths:] power
    \item[upaths:] cn\#8
    \item[upaths:] cn\#-1
    \item[sisters:] power cn\#8 cn\#-1
    \item[sisters:] times pi apply
    \item[sisters:] sin apply
    \item[sisters:] apply
  \end{description}}
  \caption{Index terms of the expression $\sin(\frac{\pi}{8})$.}
\label{fig5:encodingExample}
\end{figure*}

\subsection{Searching}

In the mathematical search system, users can input mathematical expressions using presentation MathML as a query.
The search system then uses the {\it semantic enrichment of mathematical expressions} module to convert the input expressions to Content MathML.
Figure~\ref{fig5:encodingExample2} presents an example of the terms used in the query of the expression $\sin(\frac{\pi}{8})$.
Matching is then performed using eDisMax, the default query parser of Apache Solr.
Ranking is also done using the default modified TF/IDF scores and length normalization of Apache Solr.

\begin{figure*}[tbp]
  {\ttfamily \footnotesize
  \begin{description}[font=\textbf,itemsep=0mm]
    \item[opaths:]\hfill \\ 1\#1\#apply 1\#1\#1\#sin 1\#1\#2\#apply 1\#1\#2\#1\#times 1\#1\#2\#2\#pi
1\#1\#2\#3\#apply 1\#1\#2\#3\#1\#power 1\#1\#2\#3\#2\#cn\#8 1\#1\#2\#3\#3\#cn\#-1
    \item[upaths:]\hfill \\  \#\#apply \#\#\#sin \#\#\#apply \#\#\#\#times \#\#\#\#pi \#\#\#\#apply \#\#\#\#\#power
\#\#\#\#\#cn\#8 \#\#\#\#\#cn\#-1
    \item[upaths:]\hfill \\  \#apply \#\#sin \#\#apply \#\#\#times \#\#\#pi \#\#\#apply \#\#\#\#power
\#\#\#\#cn\#8 \#\#\#\#cn\#-1
    \item[sisters:] power cn\#8 cn\#-1
    \item[sisters:] times pi apply
    \item[sisters:] sin apply
    \item[sisters:] apply
  \end{description}}
  \caption{Query terms of the expression $\sin(\frac{\pi}{8})$.}
\label{fig5:encodingExample2}
\end{figure*}

\section{Experimental Results}

\subsection{Evaluation Setup}

We collected mathematical expressions for evaluation from the Wolfram Function Site.
At the time collected, there were more than 300,000 mathematical expressions on this site.
After collection, we filtered out long expressions containing more than 20 leaf nodes to speed up the semantic enrichment because the processing time increases exponentially with the length of the expressions.
The number of mathematical expressions after filtering is approximately 20,000.
Presumably, this number is adequate for evaluating the mathematical search system.


Evaluation was done by comparing three systems:
\begin{itemize}
\item Presentation-based search with Presentation MathML (PMathML): indexing and searching are based on the Presentation MathML expressions.
\item Content-based search with semantic enrichment (SE): indexing and searching are based on the Content MathML expressions.
The Content MathML expressions are extracted automatically using semantic enrichment module.
\item Content-based search with correct Content MathML (CMathML): indexing and searching are based on the Content MathML expressions.
The Content MathML expressions are those from the Wolfram Function Site.
\end{itemize}

We used the same data to train the semantic enrichment module by 10-fold cross validation method.
The data is divided into 10 folds.
The semantic enrichment result of each fold was done by using the other 9 folds as training data.

\subsection{Evaluation Methodology}

We used ``Precision at 10'' and ``normalized Discounted Cumulative Gain'' metrics to evaluate the results.
In a large-scale search scenario, users are interested in reading the first page or the first three pages of the returned results.
``Precision at 10'' (P@10) has the advantage of not requiring the full set of relevant mathematical expressions, but its salient disadvantage is that it fails to incorporate consideration of the positions of the relevant expressions among the top $k$.
In a ranked retrieval context, normalized Discounted Cumulative Gain (nDCG) as given by Equation \ref{eq:5EQ01} is a preferred metric because it incorporates the order of the retrieved expressions.
In Equation~\ref{eq:5EQ01}, Discounted Cumulative Gain (DCG) can be calculated using the Equation~\ref{eq:5EQ02}, where $rel_{i}$ is the graded relevance of the result at position $i$.
Ideal DCG (IDCG) is calculable using the same equation, but IDCG uses the ideal result list which was sorted by relevance.

\begin{equation}
\centering
\mathrm{nDCG_{p}} = \frac{DCG_{p}}{IDCG_{p}} 
\label{eq:5EQ01}
\end{equation}

\begin{equation}
\centering
\mathrm{DCG_{p}} = rel_{1} + \sum_{i=2}^{p} \frac{rel_{i}}{\log_{2}(i)}
\label{eq:5EQ02}
\end{equation}

For performance analysis of the mathematical search system, we manually created 15 information needs (queries) and used them as input queries of our mathematical search system.
The queries are created based on NTCIR queries with minor modification. Therefore, the search system always gets at least one exact match.
Table~\ref{tab5:query} shows the queries we used.
The top 10 results of each query were marked manually as relevant ($rel=1$), non-relevant ($rel=0$), or partially relevant ($rel=0.5$).
The system then calculates P@10 and an nDCG value based on the manually marked results.

\begin{table}[htb]
\centering
\caption{Queries.}
\begin{tabular}{|L{1cm}|C{5cm}|}
\hline
\centering\textbf{No.} & \centering\textbf{Query} \tabularnewline \hline
1 & $ \int_0^\infty \! x \, \mathrm{d}x	$ \\ 
2 & $ x^2 + y^2 $ \\
3 & $ \int_0^\infty \! e^{-x^2} \, \mathrm{d}x $		\\
4 & $ arcsin(x) $										\\ 
5 & $ k^2 $												\\ 
6 & $ \frac{\cosh e z + \sinh e z}{e} $					\\ 
7 & $ \mathcal{R}_z \Psi^{\nu}(z),\tilde{\infty} $		\\
8 & $ \int \frac{a^{d+bz}}{z} \mathrm{d}z	$			\\ 
9 & $ \lim_{\nu \to \infty} \frac{L_{\alpha + \nu}}{L_{\nu}}	$			\\
10 & $ \mathcal{B} \mathcal{P}_z \mathfrak{B} ^{\mu}_{\nu}(z) $		\\
11 & $ \nu \in \mathbb{N} $		\\
12 & $ \Psi^{\nu}(z) $  \\
13 & $ \log(z+1) $  \\
14 & $ H_n(z) $  \\
15 & $ \frac{1}{\pi} \int_0^\pi (\cos t n - z\sin t) \mathrm{d}t	$ \\ 
\hline
\end{tabular}
\label{tab5:query}
\end{table}

\subsection{Experimental Results}

\begin{figure*}[htb]
\centering
\includegraphics[width=12cm]{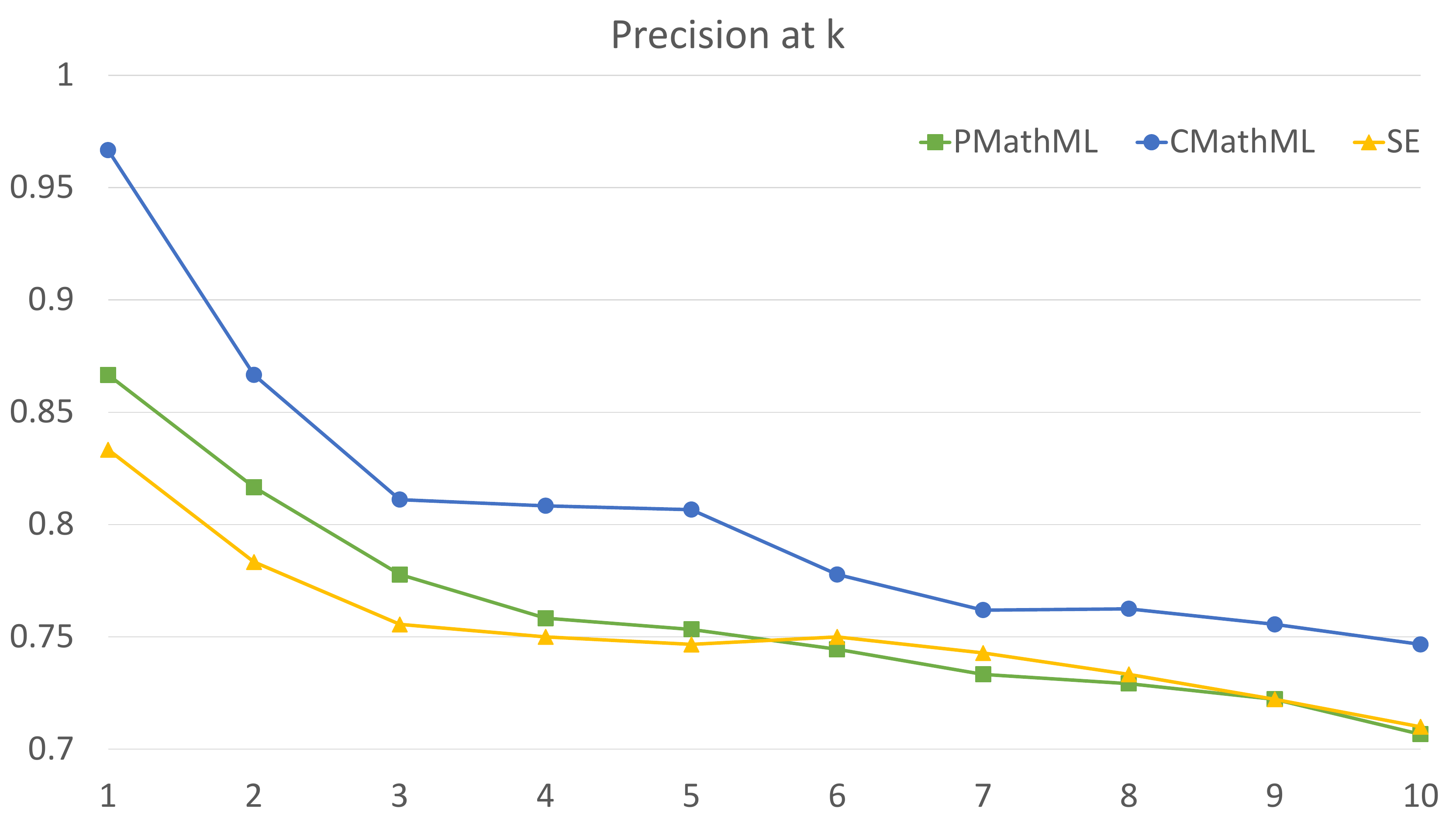}
\caption{Top 10 precision of the search system.}
\label{fig5:CON3}
\end{figure*}

\begin{figure*}[htb]
\centering
\includegraphics[width=12cm]{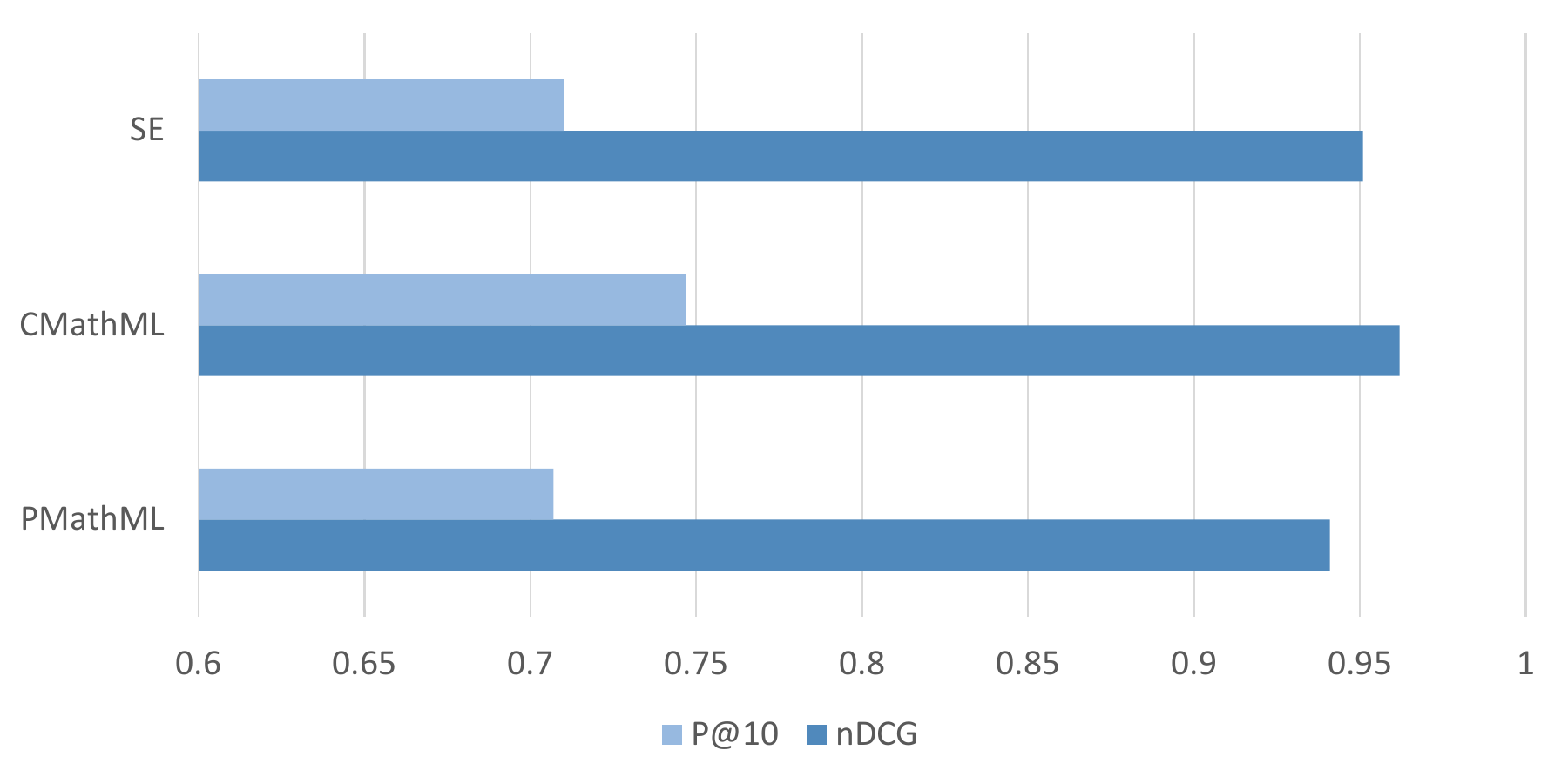}
\caption{Comparison of different systems.}
\label{fig5:COMPARE}
\end{figure*}

Comparisons among the three systems were made using P@10 and nDCG scores.
Table~\ref{tab5:result1} and figure~\ref{fig5:COMPARE} show the P@10 and nDCG scores obtained from the search.
Figure~\ref{fig5:CON3} depicts the top 10 precision of the search system.
The x axis shows the $k$ number, which ranges from 1 to 10.
The y axis shows the precision score.
The precision score decreased, while $k$ increased, which indicates that the higher results are more relevant than lower results.

\begin{table}[htb]
\centering
\caption{nDCG and P@10 scores of the search systems.}
\begin{tabular}{|L{2cm}|R{1.2cm}|R{1.2cm}|}
\hline
\centering\textbf{Method} 	& \centering\textbf{nDCG} & \centering\textbf{P@10}\tabularnewline \hline
PMathML	 	& 0.941 & 0.707 \\ 
CMathML		& 0.962 & 0.747 \\ 
SE			& 0.951 & 0.710 \\ \hline
\end{tabular}
\label{tab5:result1}
\end{table}

In the experiment, a strong relation between {\it semantic enrichment of mathematical expressions} and {\it content-based mathematical search system} was found.
As shown in Nghiem et. al~\cite{2013-IEICE-Minh}, the error rate of {\it semantic enrichment of mathematical expressions} module is around 29 percent.
With current performance, using this module for the mathematical search system still improves the search performance.
The system gained 1 percent in nDCG score and 0.3 percent in P@10 score compared to the Presentation MathML-based system.
Overall, the system using perfect Content MathML yielded the highest results.
In direct comparison using nDCG scores, the system using semantic enrichment is superior to the Presentation MathML-based system, although not by much.
Out of 15 queries, the semantic enrichment system showed better results than Presentation MathML-based system in 7 queries, especially when the mathematical symbols contain specific meanings, e.g. Poly-Gamma function (query 10), Hermite-H function (query 14).
In case the function has specific meaning but there is no ambiguity representing the function, e.g. Legendre-Q function (query 12), both systems give similar results.
Presentation MathML system, however, produced better results than semantic enrichment systems in 5 queries when dealing with elementary functions (query 2, 8, 15), logarithm (query 13), and trigonometric functions (query 6) because of its simpler representation using Presentation MathML.
One exception is the case of query 4, when there is more than one way to represent an expression with a specific meaning, e.g. $sin^{-1}$ and $arcsin$, Presentation MathML system gives unstable results.

This finding, while preliminary, suggests that we can choose either search strategy depending on the situation.
We can use Presentation MathML system for elementary functions or when there is no ambiguity in the Presentation MathML expression.
Otherwise, we can use a Content MathML system while dealing with functions that contain specific meanings.
Another situation in which we can use a Content MathML system is when there are many ways to present an expression using Presentation MathML markup.

The average time for searching for a mathematical expression is less than one second on our Xeon 32 core 2.1 GHz 32 GB RAM server.
The indexing time, however, took around one hour for 20,000 mathematical expressions.
Because of the unavailability of standard corpora to evaluate content-based mathematical search systems, the evaluation at this time is quite subjective and limited.
Although this study only uses 20,000 mathematical expressions for the evaluation, the preliminary experimentally obtained results indicated that the semantic enrichment approach showed promise for content-based mathematical expression search.

\section{Conclusion}

By using semantic information obtained from semantic enrichment of mathematical expressions system, the content-based mathematical search system has shown promising results.
The experimental results confirm that semantic information is helpful to the mathematical search.
Depending on the situation, we can choose to use either presentation-based or content-based strategy for searching.
However, this is only a first step; many important issues remain for future studies.
Considerably more work will need to be done using a larger collection of queries.
In addition, there are many other valuable features that are worth
considering besides the semantic markup of an expression, such as the description of the formula and its variables.

\section*{Acknowledgments}

This work was supported by JSPS KAKENHI Grant Numbers 2430062, 25245084.

\bibliographystyle{splncs} 

\bibliography{references}

\begin{thebibliography}{10}

\bibitem{2013-NTCIR-Aizawa}
Aizawa, A., Kohlhase, M., Ounis, I.:
\newblock {NTCIR-10 Math} pilot task overview.
\newblock In: National Institute of Informatics Testbeds and Community for
  Information access Research 10 (NTCIR-10). (2013)  654--661

\bibitem{URL-DLMF}
{National Institute of Standards and Technology}:
\newblock Digital library of mathematical functions.
\newblock \url{http://dlmf.nist.gov} ((visited on 01 March. 2014))

\bibitem{URL-SpringerLaTeXSearch}
Springer:
\newblock {Springer LaTeX Search}.
\newblock \url{http://www.latexsearch.com/} ((visited on 01 March. 2014))

\bibitem{2005-DLMF-Youssef}
Youssef, A.S.:
\newblock Information search and retrieval of mathematical contents: Issues and
  methods.
\newblock In: The ISCA 14th International Conference on Intelligent and
  Adaptive Systems and Software Engineering. (2005)  100--105

\bibitem{2008-Query-Altamimi}
Altamimi, M.E., Youssef, A.S.:
\newblock A math query language with an expanded set of wildcards.
\newblock Mathematics in Computer Science \textbf{2} (2008)  305--331

\bibitem{2007-Query-Youssef}
Youssef, A.S., Altamimi, M.E.:
\newblock An extensive math query language.
\newblock In: SEDE. (2007)  57--63

\bibitem{2007-Query-Miner}
Miner, R., Munavalli, R.:
\newblock An approach to mathematical search through query formulation and data
  normalization.
\newblock In: Towards Mechanized Mathematical Assistants. Volume 4573 of
  Lecture Notes in Computer Science.
\newblock Springer Berlin Heidelberg (2007)  342--355

\bibitem{URL-Wolfram}
Wolfram:
\newblock {The Wolfram Functions Site}.
\newblock \url{http://functions.wolfram.com/} ((visited on 01 March. 2014))

\bibitem{2008-MathGO-Adeel}
Adeel, M., Cheung, H.S., Khiyal, S.H.:
\newblock Math go! prototype of a content based mathematical formula search
  engine.
\newblock Journal of Theoretical and Applied Information Technology Vol. 4, No.
  10 (2008)  1002--1012

\bibitem{2006-MWS-Kohlhase}
Kohlhase, M., Sucan, I.A.:
\newblock A search engine for mathematical formulae.
\newblock Artificial Intelligence and Symbolic Computation Lecture Notes in
  Computer Science Vol. 4120 (2006)  241--253

\bibitem{2009-MathDa-Yokoi}
Yokoi, K., Aizawa, A.:
\newblock An approach to similarity search for mathematical expressions using
  mathml.
\newblock In: 2nd Workshop Towards a Digital Mathematics Library, DML 2009
  (2009)  27--35

\bibitem{2013-NTCIR-Kohlhase}
Kohlhase, M., Prodescu, C.C.:
\newblock Mathwebsearch at {NTCIR-10}.
\newblock In: National Institute of Informatics Testbeds and Community for
  Information access Research 10 (NTCIR-10). (2013)  675--679

\bibitem{2012-CIKM-Nguyen}
Nguyen, T.T., Chang, K., Hui, S.C.:
\newblock A math-aware search engine for math question answering system.
\newblock In: Proceedings of the 21st {ACM} international conference on
  Information and knowledge management (CIKM 2012). (2012)  724--733

\bibitem{2010-MathML}
Ausbrooks, R., Buswell, S., Carlisle, D., Chavchanidze, G., Dalmas, S., Devitt,
  S., Diaz, A., Dooley, S., Hunter, R., Ion, P.,  et~al.:
\newblock Mathematical markup language ({MathML}) version 3.0. {W3C}
  recommendation.
\newblock World Wide Web Consortium (2010)

\bibitem{2004-OpenMath}
Buswell, S., Caprotti, O., Carlisle, D.P., Dewar, M.C., Gaetano, M., Kohlhase,
  M.:
\newblock The openmath standard.
\newblock Technical report, version 2.0. The Open Math Society, 2004 (2004)

\bibitem{URL-SnuggleTeX}
McKain, D.:
\newblock {SnuggleTeX version 1.2.2}.
\newblock \url{http://www2.ph.ed.ac.uk/snuggletex/} ((visited on 01 March.
  2014))

\bibitem{URL-LaTeXML}
Miller, B.R.:
\newblock {LaTeXML} a {LaTeX} to {XML} converter.
\newblock \url{http://dlmf.nist.gov/LaTeXML/} ((visited on 01 March. 2014))

\bibitem{2009-DML}
Kamali, S., Tompa, F.W.:
\newblock Improving mathematics retrieval.
\newblock In: 2nd Workshop Towards a Digital Mathematics Library. (2009)
  37--48

\bibitem{DBLP-KamaliT13}
Kamali, S., Tompa, F.W.:
\newblock Structural similarity search for mathematics retrieval.
\newblock In: MKM/Calculemus/DML. (2013)  246--262

\bibitem{DBLP:journals/ijdar/ZanibbiB12}
Zanibbi, R., Blostein, D.:
\newblock Recognition and retrieval of mathematical expressions.
\newblock IJDAR \textbf{15} (2012)  331--357

\bibitem{2006-Mathfind-Munavalli}
Munavalli, R., Miner, R.:
\newblock Mathfind: a math-aware search engine.
\newblock In: Proceedings of the 29th annual international ACM SIGIR conference
  on Research and development in information retrieval, ACM (2006)  735--735

\bibitem{2003-DLMF-Miller}
Miller, B.R., Youssef, A.S.:
\newblock Technical aspects of the digital library of mathematical functions.
\newblock Annals of Mathematics and Artificial Intelligence \textbf{38} (2003)
  121--136

\bibitem{2007-DLMF-Youssef}
Youssef, A.S.:
\newblock Methods of relevance ranking and hit-content generation in math
  search.
\newblock In: Towards Mechanized Mathematical Assistants. (2007)  393--406

\bibitem{MIaS2011}
Sojka, P., L{\'\i}{\v s}ka, M.:
\newblock {The Art of Mathematics Retrieval}.
\newblock In: {Proceedings of the ACM Conference on Document Engineering,
  DocEng 2011}, {Mountain View, CA}, {Association of Computing Machinery}
  (2011)  57--60

\bibitem{EgoMath2011}
Mišutka, J., Galamboš, L.:
\newblock System description: Egomath2 as a tool for mathematical searching on
  wikipedia.org.
\newblock In Davenport, J., Farmer, W., Urban, J., Rabe, F., eds.: Intelligent
  Computer Mathematics. Volume 6824 of Lecture Notes in Computer Science.
\newblock Springer Berlin Heidelberg (2011)  307--309

\bibitem{URL-ActiveMath}
Siekmann, J.:
\newblock Activemath.
\newblock \url{http://www.activemath.org/eu/} ((visited on 01 March. 2014))

\bibitem{URL-MathOverflow}
MathOverflow:
\newblock Math overflow.
\newblock \url{http://mathoverflow.net/} ((visited on 01 March. 2014))

\bibitem{2013-IEICE-Minh}
Nghiem, M.Q., Kristianto, G.Y., Aizawa, A.:
\newblock Using mathml parallel markup corpora for semantic enrichment of
  mathematical expressions.
\newblock Journal of the Institute of Electronics, Information and
  Communication Engineers, vol.E96-D, no.8 \textbf{69} (2013)  1707--1715

\bibitem{URL-Solr}
Apache:
\newblock Apache solr.
\newblock \url{http://lucene.apache.org/solr/} ((visited on 01 March. 2014))

\bibitem{2013-NTCIR-Goran}
Topic, G., Kristianto, G.Y., Nghiem, M.Q., Aizawa, A.:
\newblock The {MCAT} math retrieval system for {NTCIR-10 Math} track.
\newblock In: National Institute of Informatics Testbeds and Community for
  Information access Research 10 (NTCIR-10). (2013)  680--685

\bibitem{URL-arXiv}
{Cornell University Library}:
\newblock arxiv.
\newblock \url{http://arxiv.org/} ((visited on 01 March. 2014))

\end{thebibliography}

\end{document}